# Measurement of optical second-harmonic generation from an individual single-walled carbon nanotube

M. J. Huttunen[1], O. Herranen[2], A. Johansson[2,5], H. Jiang[3], P. R. Mudimela[3], P. Myllyperkiö[4], G. Bautista[1], A. G. Nasibulin[3], E. I. Kauppinen[3], M. Ahlskog[2], M. Kauranen[1] and M. Pettersson[4]

[1] Department of Physics, Tampere University of Technology, P.O. Box 692, FI-33101 Tampere, Finland.

[2] Nanoscience Center, Department of Physics, P.O. Box 35, FI-40014 University of Jyväskylä, Finland.

[3] Department of Applied Physics and Center for New Materials, Aalto University, Puumiehenkuja 2, FI-00076 Aalto University, Finland.

[4] Nanoscience Center, Department of Chemistry, P.O. Box 35, FI-40014 University of Jyväskylä, Finland.

E-mail: andreas.johansson@jyu.fi

**Abstract.** We show that optical second-harmonic generation (SHG) can be observed from individual single-walled carbon nanotubes (SWCNTs) and, furthermore, allows imaging of individual tubes. Detailed analysis of our results suggests that the structural noncentrosymmetry, as required for SHG, arises from the non-zero chiral angle of the SWCNT. SHG thus has potential as a fast, non-destructive, and simple method for imaging of individual nanomolecules and for probing their chiral properties. Even more, it opens the possibility to optically determine the handedness of individual SWCNTs.

---

[5] Author to whom any correspondence should be addressed.

Measurement of optical second-harmonic generation from an individual single-walled carbon nanotube

**1. Introduction**

Single-walled carbon nanotubes (SWCNTs) have a rich variety of physical properties depending on very small differences in how each individual graphene layer is rolled up into a tube. The wide variety of electronic structures in combination with a mechanically strong nanoscale lattice and an exceptionally high thermal conductivity are among the main reasons for the large interest in using SWCNTs in future electronic and optical applications [1]. Unfortunately, scaled-up production of SWCNTs with a specific set of (n,m) indices is still very challenging, necessitating a lot of enrichment after SWCNTs have been synthesized [2]. The promise for new electronic and optical applications of SWCNTs is partly held up by this challenge [1]. There is thus a great need for new characterization methods that are fast, precise and cost-effective to verify the composition of SWCNTs, subsequently enabling the fabrication of devices based on the properties of individual SWCNTs.

Optical spectroscopy can, in spite of its limitations in spatial resolution, provide valuable information about nano-objects [3], including individual SWCNTs [4,5]. Single-photon excitation processes, such as Raman spectroscopy [6,7,8], electronic absorption [9], Rayleigh scattering [10] and luminescence [7,11,12] have provided information on electronic and phononic structure and their dynamics in SWCNTs. Nonlinear optical techniques, on the other hand, probe the response of the system to multiple interactions with the electromagnetic field, and can therefore provide complementary information. Nonlinear techniques are especially interesting, because nanomaterials can have remarkably large nonlinear optical responses [3]. It can even be anticipated that nonlinear spectroscopy on the single particle level will become an important characterization technique. This would be very desirable for nanomaterials, which are often structurally inhomogeneous. Indeed, four-wave mixing (FWM) in the picosecond regime has been used to image carbon nanotubes (CNTs) on a surface [13] and in the femtosecond regime to address suspended, individual SWCNTs [14]. As a third-order nonlinear process, FWM has no symmetry requirements for the system under study, thus providing a very general capability.

In contrast, second-order processes, such as second-harmonic generation (SHG), are very sensitive to symmetry, and are forbidden in centrosymmetric materials within the electric-dipole approximation of the light-matter interaction [15]. Chirality is a structural property that necessarily breaks centrosymmetry, allowing second-order processes. A vast majority of SWCNTs are chiral, and SHG has been proposed as a probe of their chirality [16-19]. Beyond chirality itself, the handedness of SWCNTs is important for applications such as molecular recognition [20,21]. Although Raman spectroscopy has been developed towards the determination of the chiral indices of SWCNTs [8], it is not able to determine the absolute handedness.

SHG has been observed from thin films of SWCNTs [17,18,22], and from an ensemble of SWCNTs trapped in zeolites [23]. The latter result was associated with chirality, and a large second-order susceptibility on the order of 400 pm·V$^{-1}$, predicted theoretically [16,24], was experimentally confirmed. It is, however, not evident *a priori* whether individual SWCNTs have a sufficient second-order response to be detected by SHG. Such SHG from an individual SWCNT could also be interpreted as the most elementary form of hyper-Rayleigh scattering, the term usually used for SHG scattering from an incoherent ensemble of molecules [25].

In this Letter, we demonstrate that not only can individual SWCNTs be observed by SHG, but also SHG imaging is possible. Our CNTs are suspended across a slit-opening in a membrane, hence providing a highly symmetric environment for the tubes. Prior to SHG measurements, the CNTs are carefully characterized by transmission electron microscopy (TEM) and electron diffraction (ED) in order to confirm their individuality and determine their chiral indices (n,m). In spite of having an individual SWCNT with relatively low chiral angle, its SHG image is observed. The results demonstrate the potential of SHG as a non-destructive tool to study individual SWCNTs and their symmetry properties.

**2. Experiment**

*2.1. Sample preparation*

Our samples consist of SWCNTs grown with chemical vapor deposition (CVD) across a 1.2 µm wide slit opening in a 300 nm thick Si$_3$N$_4$ membrane. A schematic of the sample geometry is shown in figure 1.

The sample preparation starts from a 500 $\mu$m thick, double side polished <100> Si wafer with a 300 nm thick dielectric layer of $Si_3N_4$ on both sides. First, a 750 $\mu$m x 750 $\mu$m opening in the bottom $Si_3N_4$ layer was made, using optical lithography followed by reactive ion etching (RIE) at a pressure of 55 mTorr and 150 W power with a gas flow of 50 sccm of $CHF_3$ and 5 sccm of $O_2$. This was followed by wet etching through the Si wafer in 35 % KOH at 97 ° C, with an etch rate of 180 $\mu$m·h$^{-1}$. The etching process is anisotropic with an etching angle of 54.7 °, resulting in a 50 $\mu$m x 50 $\mu$m $Si_3N_4$ membrane window on the front side. A slit opening with the dimensions 1.2 $\mu$m x 40 $\mu$m was then made in the $Si_3N_4$ membrane, using electron beam lithography followed by a second RIE step. The bottom of the sample was then covered by a 25 nm thick layer of tantalum, chosen for its high melting point, which enables it to survive during the CVD of CNTs. The metal layer supports the membrane and can be used for gating purposes. SWCNTs were grown across the membrane in a vertical CVD reactor. CO was used as the carbon source, Ni as the catalyst material and the CNT growth temperature was 750 °C. The SWCNT synthesis is described in more detail elsewhere [26,27].

*2.2. TEM and ED characterization*

The morphology and structure of the suspended CNTs were characterized prior to SHG imaging in a Philips CM200-FEG TEM, which is equipped with a Gatan 794 multiscan charge-coupled device camera for digital data recording. Both direct imaging and ED patterning were used in several locations along the nanotube bridges to study their composition and chiral properties. To minimize the risk of electron beam induced damage to the CNT lattice, a low acceleration voltage of 80 kV and short exposure times of 3 seconds were used. Thus the electrons in the beam carried significantly lower energy than the threshold of about 100 keV for knock-on damage to remove a carbon atom from the CNT lattice [28, 29].

For the sample in figure 2a, CNTs were found at two locations, creating nanometer-wide bridges across the slit opening. ED patterning revealed that the nano-bridge on the left consists of a bundle of 4-5 CNTs. The nano-bridge on the right was confirmed to be an individual SWCNT, except at the almost horizontal section close to the upper edge, where at least one other short CNT is bundled to it. The ED pattern from the individual SWCNT is shown in figure 2b, from which the chiral indices of the SWCNT were determined to be (42,1), using intrinsic layer line distance analysis [30]. The chiral indices denote the number of lattice base vectors needed to describe the roll-up vector, and thereby define the lattice structure of the SWCNT. In the present case the nanotube has a diameter of 3.3 nm and a chiral angle (i.e., the smallest angle between the tube cross-section and a lattice base vector) of 1.2 degrees. The appearance of separated layer-lines in the ED pattern (arrows in figure 2b) confirms that the SWCNT is chiral [31]. The TEM image shown in figure 2a was taken after the SHG measurement in a JEOL JEM-2200FS microscope.

*2.3. SHG microscopy*

The SHG microscope setup is schematically shown in figure 1. The light source was a mode-locked Nd:Glass laser emitting an 82-MHz train of 200 fs pulses centered at the wavelength of 1060 nm. The average power of the pulsed laser beam was around 1 mW to avoid sample damage. The input beam was expanded to a diameter of 7 mm, spatially filtered, well collimated and polarized before entering the focusing objective (NA=0.8).

The structure and non-vanishing components of the second-order susceptibility tensor depend on material symmetry, giving rise to polarization-dependent SHG response. In order to facilitate coupling of the fundamental laser beam with all the possible components of the tensor and for any orientation of the SWCNTs, a calcite Glan polarizer was used to clean up the linear polarization of the beam and a subsequent quarter-wave plate was used to change the polarization to circular. SHG detection on the other hand was unpolarized to collect as much light as possible with no accidental discrimination.

The SHG light was collected only in reflection by the focusing objective and separated from the fundamental and possible two-photon fluorescent light by consecutive long-pass dichroic mirror and narrowband interference filter (16.5 nm bandwidth centered at 532 nm), and detected by a cooled photomultiplier tube connected to a photon counting unit. To ease the sample positioning, a white-light imaging arm was implemented in the microscope. To avoid changes to the input polarization of the laser beam due to the imaging arm, a flip mirror was used to steer the white light to the imaging lens and a consecutive camera when needed.

Imaging was done by raster scanning the sample at the focal plane of the microscope objective using a 3-axis piezo-actuated translation stage. The pixel dwell time was 150 ms, averaged twice, and for the 5 mm x 5 mm scanning area, 100 x 100 pixels were used.

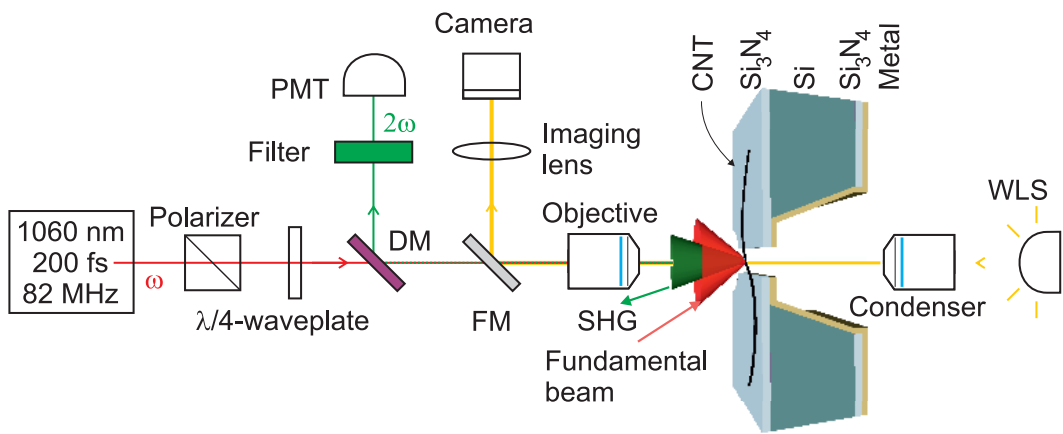

**Figure 1.** Schematics of sample geometry and SHG microscopy setup. The SWCNTs are suspended across a 1.2 mm wide slit opening in a 300 nm thick $Si_3N_4$ membrane. A pulsed and collimated beam from an ultrafast laser was passed through a polarizer and a quarter-wave plate to generate circularly polarized light and focused to the sample by an objective (NA=0.8). Reflected SHG (2w) was separated from the fundamental excitation wavelength (w) by a dichroic mirror (DM) and an interference filter and detected by a photomultiplier tube (PMT). A white-light source (WLS), flip mirror (FM) and bright-field imaging arm were used for alignment of the sample for SHG measurements.

## 3. Results

The SHG image is shown in figure 2c, clearly depicting the SWCNTs suspended over the slit. Note that the strong background signal from $Si_3N_4$ prevented SHG visualization of SWCNTs on top of the membrane. The signal-to-noise ratio (SNR) for the suspended part of the SWCNT was estimated from a line profile, formed by averaging 8 adjacent line profiles to be ~2 (shown in figure 2d, with its location as a dotted line in figure 2c). While the SNR is small, the features are distinct and correspond well to the TEM image.

In order to verify that the data is not coincidental and specific to this sample only, we performed measurements on additional samples and observed similar data from 8 different CNTs. Although the individuality and chiral indices of these additional CNTs were not determined, the SHG signals always correlated well with the TEM or scanning electron microscope (SEM) images. Figure 3 shows SEM and SHG images taken from a sample with several nanotube bridges, marked by roman numerals I-V. Interestingly, not all the nanotube bridges show up in the SHG image. This could imply that resonance enhancement plays an important role in SHG from the nanotubes. Another possible explanation is that the dark tubes are achiral and thus not SHG-active. In addition, we observed that for some of the nanotubes the SHG signal degrades with time with no sign of recovery. The reason for signal degradation from some of the nanotubes is not fully clear, as the CNTs after SHG measurements are confirmed still present by TEM or SEM imaging.

Due to the weakness of the SHG signal, which rapidly vanishes below the noise level as the excitation power is reduced, it is difficult to reliably measure its spectrum or verify its quadratic dependence on the fundamental intensity. Furthermore, the quadratic dependence is characteristic to any two-photon-excited process and, as such, does not provide evidence of the SHG origin of the signal. We therefore carefully provide estimates for other possible origins of the signal and successfully exclude them. We first estimate the expected SHG signal strength by assuming that the second-order susceptibility of CNTs is equal to 400 pm·V$^{-1}$, as reported in Ref. [23]. This value is more than a factor of 100 larger than the typical scale of second-order susceptibility for conventional materials. In consequence, our modest 1 mW average laser power and $6.4 \times 10^{-13}$ m$^2$ focused spot size, should give rise to a very large SHG source polarization of $13 \times 10^{-5}$ W·s·V$^{-1}$·m$^{-2}$, yielding a source dipole moment of $9.2 \times 10^{-28}$ W·m·s·V$^{-1}$ within the illuminated volume. By neglecting field polarization effects, this dipole moment emits a total power of $9 \times 10^{-8}$ W into all solid angles. The light collection efficiency of the objective is ~$10^{-1}$ and the efficiency of the detection channel ~$10^{-1}$, yielding the estimated SHG signal count rate of $4 \times 10^4$ s$^{-1}$. This is about 500 times higher than the measured signal of 77 counts·s$^{-1}$. Note, however, that the signal scales quadratically with the susceptibility and 400 pm·V$^{-1}$ is an exceptionally high value. The difference between the experiment and estimate could therefore be due to the fact that the susceptibility of our particular tube is somewhat lower than that in Ref. [23], yielding about 18 pm·V$^{-1}$ for our particular tube.

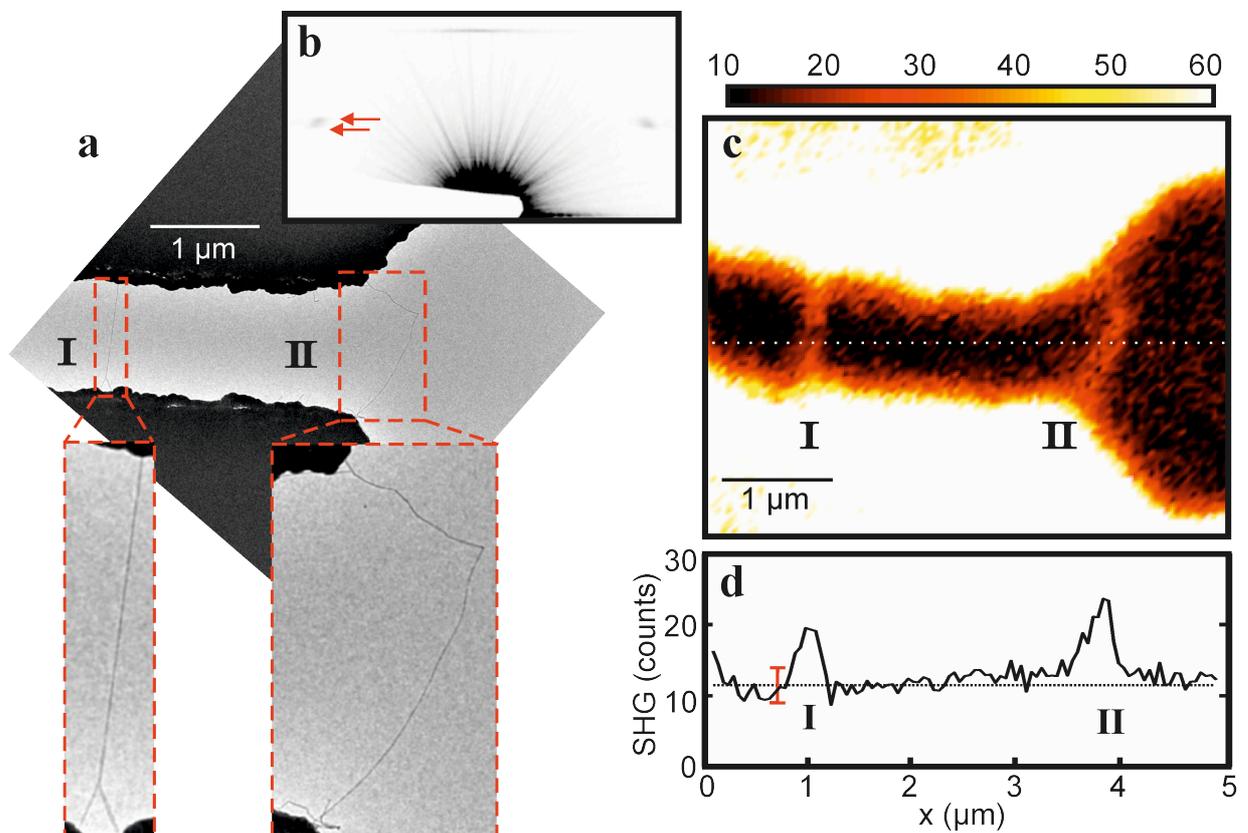

**Figure 2.** TEM and SHG characterization of suspended SWCNTs. (**a**) TEM image of the sample, featuring a bundle of 4-5 CNTs on the left (I) and an individual SWCNT on the right (II). (**b**) ED pattern from (II). The separated layer-lines indicated by two arrows clearly demonstrate that the nanotube has a chiral structure, which is very close to a zigzag tube. (**c**) SHG image of the sample showing the two suspended SWCNT structures (I and II) over the air slit. (**d**) A line profile, formed by averaging 8 adjacent line profiles, was taken from the SHG image (dotted white line in (**c**)) to visualize the SNR of the measurement. The mean background value (11.5 counts) of the measurement is shown as a dotted line and the calculated standard deviation (5 counts) as a red bar.

Measurement of optical second-harmonic generation from an individual single-walled carbon nanotube

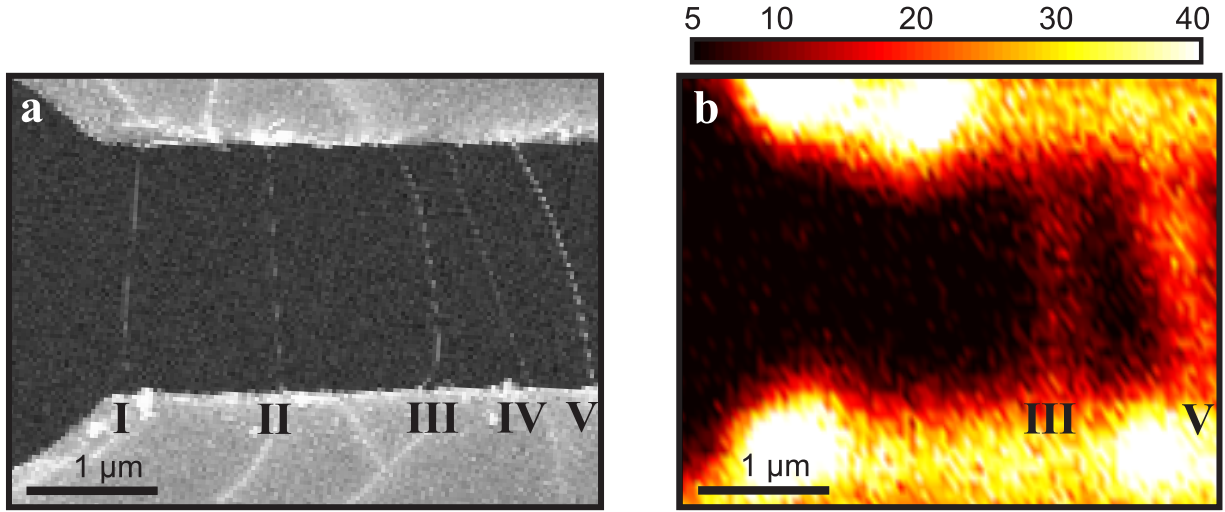

**Figure 3.** (a) SEM and (b) SHG images of SWCNTs suspended across the horizontal slit opening (dark) in the $Si_3N_4$ membrane (white area). The Roman numerals mark the position of SWCNT nano-bridges. Interestingly, only some of the nano-bridges are visible in the SHG image.

To exclude scattering of fundamental light into our detector as a source of the signal, we assume that the linear susceptibility of the CNT is on the order of unity [32]. By a similar calculation, we obtain a source dipole moment of $8.5 \times 10^{-27}$ W·m·s·V$^{-1}$, which should lead to a signal of $4 \times 10^7$ counts·s$^{-1}$ in agreement with Ref. [32]. However, our setup has a strong discrimination between the fundamental and SHG wavelengths from the dichroic mirror (>10), from the interference filter (>$10^5$), and from the detector efficiency (>$10^3$). The discrimination is therefore at least nine orders of magnitude, and most likely much higher. The scattered signal is thus at most $4 \times 10^{-2}$ s$^{-1}$, much less than the detected signal. To verify this and show that the measured SHG signal is due to nonlinear interaction between the fundamental beam and the SWCNTs, we replaced the SWCNT sample with a strong scatterer (reflecting substrate). The signal then vanished completely, and did not reappear even after 20-fold increase in the input laser power thereby excluding scattering of fundamental light or existing SHG from preceding components as the origin of the signal.

We next consider linear and multiphoton absorption induced luminescence. The linear absorption cross section of SWCNTs is very high, such as $10^{-17}$ cm$^2$/carbon atom for (6,5) tube [33]. This would lead to very high absorption rate under our conditions. However, the luminescence saturates at very low excitation numbers of 2–6 excitons per pulse for air-suspended SWCNTs with a length of 2–5 μm [34]. Using the value of 2 excitons per pulse, the excitation rate is $1.6 \times 10^8$ s$^{-1}$. Using 20 % for the emission quantum yield, and taking into account the detection and collection efficiencies ($0.1 \times 0.1$) gives $3.2 \times 10^5$ photons s$^{-1}$. Taking into account the filtering and detector efficiency (factor of $10^9$) the maximal count rate due to linear absorption induced photoluminescence is $3.2 \times 10^{-4}$ s$^{-1}$. Thus, it cannot be the source of the observed signal.

For two-photon absorption, an absorption coefficient of $5.4 \times 10^{-7}$ cm·W$^{-1}$ was obtained for a homogeneous film of HiPCO SWCNTs [35]. Under our conditions, this would lead to higher rate of absorption than the saturation limit of 2 excitons per pulse for photoluminescence. Thus, we again start from the above excitation level of $1.6 \times 10^8$ s$^{-1}$ to estimate two-photon-induced luminescence. The luminescence which occurs after relaxation of the system to the bottom of the band ($E_{11}$ transition) can be neglected because the emission for a thick tube occurs in the mid-infrared, which is not detectable by the setup. However, we have to consider the very fast non-radiative relaxation and the fraction of light that can be emitted while the system is still energetically in the detection window of the SHG filter. Upon excitation of the $E_{22}$ transition, intraband relaxation of electrons and holes occurs fully within 100 fs [36], i.e., the relaxation lifetime is much shorter than this. Thus, we estimate the nonradiative relaxation rate as $1 \times 10^{14}$ s$^{-1}$. If we assume the radiative relaxation rate as $1 \times 10^9$ s$^{-1}$, the quantum yield for emission is $1 \times 10^{-5}$ giving the emission rate of $1.6 \times 10^3$ s$^{-1}$. This value corresponds to hot luminescence, i.e., emission of a non-relaxed system. Hot luminescence is emitted over broad spectral range which can be taken as the energy difference between the excitation energy and the $E_{11}$ energy, which is ~2 eV [37]. The SHG filter bandpass is 0.072 eV wide, which transmits a fraction of 0.036 of the total hot luminescence. Taking this and the detection and collection efficiencies (0.01) into account yields a count rate of 0.6 s$^{-1}$. This is already two orders of magnitude lower than the observed rate but there is one more factor to consider. Two-photon excitation obeys different selection rules than one-photon excitation [38] and accesses even parity states (2p, 3p, ...), which are higher in energy than the lowest one-photon accessible state (1s). The energy difference between the 2p and 1s states is ~300 meV for several tubes [38]. Thus, fluorescence is expected to occur at significantly lower energies than two-photon excitation as has been observed also experimentally [38]. Altogether, we can safely exclude two-photon induced luminescence as a possible source of the observed signal.

To our knowledge, the three-photon absorption cross-sections of SWCNTs are not known. For multi-walled nanotubes, the three-photon absorption coefficient γ is 1.36 × 10$^{-20}$ cm$^6$W$^{-2}$ [39]. This and the thickness of 3.3 nm of our tube yields a value of 4 x 10$^{-7}$ for the relative change in intensity due to three-photon absorption. Taking into account the ratio between the area of the nanotube and the area of the focal spot, which is 4.6 × 10$^{-3}$, the total fraction of absorbed photons is 1.8 × 10$^{-9}$. As three photons are required for absorption, the excitation rate is 3.2 × 10$^6$ s$^{-1}$. This corresponds to 0.04 excitations per pulse on average. Using the same quantum yield for hot luminescence as in the two-photon case (1 x 10$^{-5}$) yields photoluminescence rate of 32 s$^{-1}$. Again counting for the band pass filter and collection efficiencies yields a detected count rate of 0.01 s$^{-1}$. This value is well on the safe side even if the true three-photon absorption coefficient is larger than used in our estimate. All the estimations above show that the most probable explanation for the observed signal is indeed SHG.

Within the electric-dipole approximation, there are four possible sources of the SHG signal: chirality, lattice defects, tube deformations, and surface interactions. We discard the latter two, since the signal intensity is quite similar and homogeneous and is not visibly affected by tube bending either. Thus, we are left with chirality or defects (or both) as the source of SHG. The observation of a distinct ED pattern establishes that the tube is chiral and indicates a high long-range order, i.e., low defect concentration, making therefore chirality the most plausible source.

The SHG response may be enhanced by resonant transitions [23]. Using an empirical relation between the tube structure and electronic resonance energies [11], we estimate that $E_{11}$ for our (42,1) tube is 0.34 eV. Sfeir *et al.* have measured several excitonic transitions for thick tubes and obtained transition energy relations for $E_{ii}/E_{11}$ (i = 2,3,4) [40]. Using their relations for the (42,1) tube gives 0.60, 1.19 and 1.45 eV for $E_{22}$, $E_{33}$ and $E_{44}$, respectively. Another empirical relation based on a large set of Rayleigh scattering measurements gives for the same transitions similar estimates, 0.56, 1.14 and 1.50 eV, respectively [37]. From these, $E_{33}$ is close to our fundamental excitation energy of 1.17 eV (wavelength 1060 nm), suggesting that one-photon resonance is possible. For the second-harmonic excitation at 2.34 eV, only the latter empirical formula is able to give an estimate of nearby transitions [37], returning $E_{77}$ with energy of 2.40 eV as the closest transition. On the other hand, we did not observe any visible Raman signal from our SWCNTs when using an input wavelength of 532 nm, suggesting that there is no two-photon resonance in our case.

The likely chiral origin of the SHG signal provides exciting opportunities for future development. It is possible to enrich a given SWCNT handedness, e.g. by molecular recognition in combination with ultracentrifugation [20,21]. So far, the only way to confirm the handedness of a specific individual CNT is by scanning probe microscopy [41,42], or by high-resolution TEM imaging [43-45]. Both methods require vacuum environment, and are time consuming and potentially damaging to the CNT. SHG microscopy offers a much faster and non-destructive technique, and its sensitivity to chirality has been well-established [46-52].

In order to develop SHG microscopy as a probe of absolute handedness for SWCNTs, additional work is needed. The components of the second-order susceptibility can be classified as achiral or chiral. The former are non-vanishing for any noncentrosymmetric system, whereas the latter rely on chirality. In addition, the chiral components reverse their sign between the two handednesses of the chiral sample. The chiral signatures in SHG responses arise from a complex interplay between achiral and chiral susceptibility components leading to different SHG signals [53-55] depending on the handedness of the sample. The defect-free SWCNTs possess only chiral components, which will only lead to changes in the phase of the SHG signal depending on the handedness. Phase sensitive SHG techniques should therefore be used in order to determine the absolute sign of the handedness [56,57]. Alternatively, if the SWCNTs are lying on a carefully selected surface, the surface may provide the necessary achiral mixing signal for detecting a difference in the signal for SWCNTs with different handedness. It will also be important to increase the SNR for handedness measurements. For such measurements, single point SHG, instead of imaging, is sufficient but SNR needs to be improved, e.g. by increasing the measurement time from the present 150 ms per pixel.

SHG can also occur in centrosymmetric materials when interactions beyond electric dipoles (magnetic and quadrupole interactions) are taken into account. Such higher-multipole effects are expected to be significantly weaker than electric-dipole effects [58], but would allow SHG from achiral SWCNTs. To fully justify SHG as a probe of chirality of SWCNTs it will be important to study the origin of SHG from individual SWCNTs in more detail.

While we conclude that defects are not likely to explain our observations, SHG does have additional opportunities in imaging of defects. An ideal SWCNT has very high symmetry, and defects will decrease the overall symmetry and introduce new nonvanishing susceptibility components, which change the characteristics of the local SHG response. Due to its sensitivity to symmetry breaking, SHG could therefore provide a complementary technique for non-destructive and label-free visualization of defects in SWCNTs.

## 4. Summary

To conclude, we have made to our knowledge the first observation of SHG from an individual, air-suspended SWCNT. The SHG signal most likely originates from the breaking of centrosymmetry of the SWCNT by its non-zero chiral angle. We propose that SHG could be used to measure the handedness of SWCNTs and visualize defects in SWCNTs, providing a valuable tool for the study and characterization of SWCNTs. Finally, SHG techniques are fully compatible with existing nonlinear optical techniques, such as FWM, and could in the future easily be implemented for multimodal measurements of nanomolecules.

**Acknowledgements**

This study was supported by the Academy of Finland (project no. 122620, 128445, 134973, 135062 and 7122008) and CNB-E project of Aalto University MIDE program for funding. In addition, M. H. acknowledges support from Modern School of Optics and Photonics in Finland, A. J. and M. H. from Emil Aaltonen foundation, and O. H. from Jenny and Antti Wihuri foundation.

Measurement of optical second-harmonic generation from an individual single-walled carbon nanotube